\documentstyle  [11pt] {article}
\title {\bf Oscillatory behaviour in a lattice prey-predator system \vskip5mm}   
\author {{\sc Adam Lipowski}\\
\noalign{\vskip5mm}
{\it Department of Mathematics, Heriot-Watt University},\\ 
{\it EH14 4AS Edinburgh, United Kingdom}\\
and\\
{\it Department of Physics, A.~Mickiewicz University,}\\
{\it  61-614 Pozna\'{n}, Poland}\\
\noalign{\vskip5mm}}
\date {}
\newif\ifetykiety

\def\etykieta#1{\ifetykiety \par\marginpar{\centering[#1]} \fi}
\def\bibetykieta#1{\ifetykiety \marginpar{\renewcommand{\baselinestretch}{0.9}
                   \raggedright\small[#1]} \fi}

\newcommand {\SECTION} [2] {\section{#2} \label{#1} \etykieta{#1} \setcounter 
  {equation} {0}}
\newcommand {\SUBSECTION} [2] {\subsection{#2} \label{#1} \etykieta{#1} }

\newcommand {\eq} [1] {(\ref {#1})}
\newcommand {\beq} {\begin {equation}}
\newcommand {\eeq} [1] {\label {#1} \end {equation} \etykieta{#1}}
\newcommand {\beqn} {\begin {eqnarray}}
\newcommand {\eeqn} [1] {\label {#1} \end {eqnarray} \etykieta{#1}}
\catcode `\@=11
\def\@cite#1#2{#1\if@tempswa , #2\fi}
\catcode `\@=12
\newcommand {\cyt} [1] {$^{\mbox {\footnotesize \cite{#1})}}$}

\def\bib#1#2\par{\bibitem{#1} #2 \bibetykieta{#1}}
\newcommand {\fig} [1] {Fig.~\ref{#1}}
\newcounter {fig}
\newenvironment {figure_captions} {\newpage \thispagestyle {empty} \section*
{Figure captions} \begin {list} {\bf Fig.~\arabic{fig}:} {\usecounter{fig}
\settowidth{\labelwidth}{Fig.~999:} }}{\end{list}}
\def\elem#1#2\par{\item#2\label{#1}\etykieta{#1}}

\renewcommand {\baselinestretch} {1.0}
\renewcommand {\cyt} [1] {{\mbox [\cite{#1}]}}
\begin {document}
\maketitle
\begin {abstract}
Using Monte Carlo simulations we study a lattice model of a prey-predator system.
We show that in the three-dimensional model populations of preys and predators
exhibit coherent periodic oscillations but such a behaviour is absent in lower-dimensional models.
Finite-size analysis indicate that amplitude of these oscillations is finite even in the
thermodynamic limit.
In our opinion, this is the first example of a microscopic model with stochastic dynamics which
exhibits oscillatory behaviour without any external driving force.
We suggest that oscillations in our model are induced by some kind of stochastic resonance.
\end {abstract}
\SECTION{Intro}{Introduction}
Oscillatory behaviour in spatially extended systems, which appears in various forms in many
branches of physics, is still not fully understood~\cyt{NICOLIS}.
As an example of such a behaviour we can mention periodic oscillations in certain autocatalytic
reactions~\cyt{PRIG}.
From the theoretical point of view, the main problem is that inevitable fluctuations should wipe
out any coherent behaviour in such systems, thus questionning the very existence of periodic
oscillations.
Indeed, numerical analysis of certain one-dimensional reaction diffusion model ('Brusselator') 
confirms a very strong destructive role of fluctuations in such systems~\cyt{BARAS}.

Another example of this kind is the oscillatory behaviour in prey-predator systems, which is one of
the classical problems in population dynamics.
In the most transparent way such oscillations were observed for populations of hares and 
lynxes~\cyt{TANNER}.
The earliest explanation of oscillations in such systems was proposed by Lotka and
Volterra~\cyt{LOTKA}. 
In their model, populations of preys and predators are described by the following set of
differential equations
\beq
\frac{dx}{dt} = ax-bxy, \ \ \ \frac{dy}{dt} = -cy+dxy,
\eeq{1}
where $x$ and $y$ denote the number of preys and predators, respectively, and $a,b,c,d$ are
certain positive constants.

Simple analyses of model~\eq{1} indeed reveal the existence of a limit cycle, i.e.,
populations of preys and predators exhibit periodic (in time) oscillations.
However, model~\eq{1} has certain drawbacks.
In particular, it predicts an unbounded, exponential growth of the number of preys in the absence
of predators ($y=0$).
To cure this defect one has to introduce additional terms into these equations (environmental
capacity) and such terms in general destroy the limit-cycle solutions and asymptotically (i.e.,
for $t\rightarrow \infty$) the constant solutions are obtained~\cyt{SIGMUND,GOEL}.
With this respect model~\eq{1} might be more precisely termed as structurally unstable. 

In principle, one can replace right-hand sides of the above equations by more complicated functions
of $x$ and $y$, and the resulting equations ~\cyt{HALLAM} will exhibit both a limit-cycle behaviour
and remain bounded for $y=0$.
It is not clear, however, how these particular functions should be related with characteristics of 
the populations.

Recently, a lattice model of a prey-predator system was introduced~\cyt{LIPLIP}.
It was shown that in the steady state this model has two phases: (i) active phase
with a positive fraction of both preys and predators and (ii) the absorbing phase with predators
being extinct and preys invading the whole system.
For certain value of a control parameter the model undergoes a phase transition of the
directed-percolation universality class, which is actually an anticipated property, taking into
account the existence of a single absorbing state in the model's dynamics~\cyt{JANSSEN}.

An important feature of such a lattice model is that its properties might be studied using 
controllable techniques, e.g., Monte Carlo simulations, rather than postulated equations.  
Moreover, such a microscopic model takes into account fluctuations in the system which are
completely neglected in models based on differential equations like~\eq{1}.
And it is these fluctuations which are responsible for the appearance of the phase transition in
this model since the mean-field approximation, which is equivalent to a certain set of differential
equations similar to~\eq{1} (and thus neglects fluctuations), predicts that the active phase is the
generic phase of the model for all values of the control parameter and no transition takes place.

In the present paper we examine the time evolution of densities of preys $x(t)$ and predators
$y(t)$ in the above described lattice model.
One might expect that fluctuations, which in our model are caused by the stochastic nature of the
dynamics, result in a random and noncorrelated evolution of these densities.
And indeed such a behaviour is observed but only in a one-dimensional version of our model.
In two-dimensional model the behaviour of these densities is still irregular but a pronounced peak
in a Fourier transform of $x(t)$ and $y(t)$ appears and for the three-dimensional model very
regular periodic oscillations are observed.
We argue that these oscillations are induced by certain kind of stochastic resonance~\cyt{HANGI} 
and we suggest an analogy with a certain low-dimensional dynamical system examined some time ago by
Gang et al.~\cyt{GANG}.

In addition to offering a model of prey-predator systems, our results are also of some more general
interest.
They show that in spatially extended systems, intrinsic fluctuations alone might induce periodic 
oscillations.
This should be contrasted with the standard stochastic resonance setting, where
some sort of external periodic perturbation is required.

The paper is organised as follows.
In section 2 we introduce the model and only briefly describe its steady-state properties which
were already described in more detail elsewhere~\cyt{LIPLIP}.
In section 3 we present time evolution and spectral analysis of density of preys for one-, two-,
and three-dimensional version of our model.
Section 4 contains the analysis of the standard deviation of density of preys as a function of
time.
In this section we also suggest a relation with stochastic resonance.
Section 5 contains our conclusions.

\SECTION{model}{Model and its steady-state properties}
In our model a site of a $d$-dimensional cartesian lattice of linear size $L$ can be empty,
occupied by a single prey, occupied by a single predator or occupied by a single prey and a single
predator.
Dynamics of our model is specified as follows:

\noindent
(i) Choose a site at random.

\noindent
(ii) With the probability $r$ ($0<r<1$) update a prey at the chosen site (if there is one,
otherwise do nothing).
Provided that at least one neighbour of the chosen site is not occupied by a prey, the
prey (which is to be updated) produces one offspring and places it on the empty neighbouring site
(if there are more empty sites, one of them is chosen randomly).
Otherwise (i.e., when there is a prey on each neighbouring site) the prey does not breed (due to
overcrowding).

\noindent
(iii) With the probability $1-r$ update a predator at the chosen site (if there is one).
Provided that the chosen site is not occupied by a prey, the predator dies (of hunger).
If there is a prey on that site, the predator survives and consumes the prey from the site it
occupies.
If there is at least one neighbouring site which is not occupied by a predator, the predator
produces one offspring and places it on the empty site (chosen randomly when there are more such
sites).

Steady-state description of our model is given in terms of densities of preys $x$ and
predators $y$, which might be also regarded as the probabilities that a given site is occupied by a
prey or a predator, respectively.
Monte Carlo simulations of the above model predict~\cyt{LIPLIP} that this model undergoes the phase
transition at a certain value of the parameter $r=r_{{\rm c}}(d)$.
The transition point $r_{{\rm c}}(d)$ separates the active phase with $0<x,y<1$ and the absorbing
phase with $x=1, y=0$.
The plot of the steady-state densities $x$ and $y$ as functions of $r$ for the one- and
three-dimensional models, based on previous simulations~\cyt{LIPLIP}, is shown in~\fig{f1}.
Results for the two-dimensional model are not shown but they interpolate between the one- and
three-dimensional graphs with the critical point located at $r=r_{{\rm c}}(2)\sim 0.11$.
\SECTION{TIME}{Time evolution and spectral analysis}
Let us ask the following question: What is the time evolution of densities $x(t)$ and $y(t)$ in
the active phase of our model?
Because the model is driven by stochastic dynamics, the expected answer to this question is that
these quantities exhibit more or less random fluctuations.
Presented below results obtained using Monte Carlo simulations show that these expectations are
not always correct.
\SUBSECTION{d1}{d=1}
Such random fluctuations are clearly observed for the one-dimensional model in the entire active
phase (i.e., for $1>r>r_{{\rm c}}(1)\sim 0.491$) and an example for $r=0.6$ is shown in~\fig{f2}.
To analyse the time evolution more quantitatively, we calculated the Fourier power spectrum of
$x(t)$ and $y(t)$ and the results for $S(\omega)=|x(\omega)|$ are shown in~\fig{f3}.
The spectrum of $y(t)$ is similar to that of $x(t)$ and is not shown.
One can see that the spectrum is very broad, which is in agreement with a rather random pattern
observed in~\fig{f2}.
The spectrum is calculated using the intervals of 500 Monte Carlo steps and averaging is made over
100 such intervals.
The $S(\omega=0)$ value is not shown.
\SUBSECTION{d2}{d=2}
In this case populations of preys and predators also evolve in time rather irregularly 
(see~\fig{time2}).
Such irregular behaviour is reflected in~\fig{f4}, which shows that the spectrum
in this case is also very broad.
However, in a certain range of $r$ one can see a pronounced peak in the spectrum at a certain
$r$-dependent frequency.
This peak is related with the appearance of a certain slow mode which can also be seen
in~\fig{time2}.
Let us also notice that upon approaching $r_{{\rm c}}(2)(\sim 0.11)$ this peak diminishes and
shifts toward lower frequencies.
As will be shown below, the behaviour of the two-dimensional model is in some sense intermediate
between the behaviour of the one- and three-dimensional models.
\SUBSECTION{d3}{d=3}
The most interesting results are obtained for the three-dimensional model.
In~\fig{f5} we show the time evolution of $x(t)$ and $y(t)$ for $r=0.3$.
For this value of $r$ the system exhibits very regular oscillations and the spectrum
(\fig{f6}) has a very high and sharp peak.
Such regular oscillations appear only in certain range of $r$.
For sufficiently large or sufficiently  small $r$ the irregular behaviour, similar to that shown
in~\fig{f2}, sets in.
\SECTION{deviation}{Standard deviation and its finite-size analysis}
The results shown in the previous section clearly indicate a qualitative difference in temporal
evolution of one- and three-dimensional model.
Pronounced oscillations observed in the three-dimensional case prompted us to ask the following
question: What is the amplitude of these oscillations in the thermodynamic limit 
($L\rightarrow\infty$)?
To answer this question we calculated the standard deviation $\sigma$ of $x(t)$ for $d=1,2,3$ and
various system sizes $L$ and values of $r$.
This quantity roughly corresponds to the amplitude of oscillations (or fluctuations) of $x(t)$.
The behaviour of $\sigma$ as a function of $r$ is shown in~\fig{f7}.
For $d=1$ the standard deviation $\sigma$ is only weakly $r$-dependent and is a decreasing function
of $r$.
The increase of $\sigma$ upon decreasing $r$ is an expected behaviour since largest fluctuations
usually occurs at the critical point $r_{{\rm c}}(1)(\sim 0.49)$.
For the two-dimensional case $\sigma$ is also weakly $r$-dependent but one can see a small maximum
of $\sigma$ around $r=0.3$.
On the other hand, for the $d=3$ case a pronounced maximum around $r=0.3$ is observed.
Let us notice that this maximum is not related with the critical point which in this case is
located at much smaller value of $r$, namely at $r=r_{{\rm c}}(3)\sim 0.05$.
On general grounds one expects that outside critical point correlation length is finite in our
model and thus the standard deviation of $x(t)$ (and also of $y(t)$) should scale as $1/L^{d/2}$.
Thus, in the thermodynamic limit $\sigma$ should converge to zero and so should the amplitude of
oscillations.

Finite-size data which we present in~\fig{f8}-\fig{f9} show that in the three-dimensional
case this argument is false.
In~\fig{f8} we plot the standard deviation $\sigma$ as a function of $\frac{1}{L^{d/2}}$.
If the above argument about the asymptotic scaling of $\sigma$ were correct than $\sigma$ should
linearly approach zero for $L\rightarrow \infty$.
Our data show that this is indeed the case for $d=1,2$ and we expect that for $d=1,2$ such a
scaling holds for arbitrary $r$ in the active phase.
However, the behaviour for $d=3$ is different.
Although for $r=0.5$ the scaling seems to hold, it is clearly violated for $r=0.3$ where $\sigma$
does not even converge to zero. 
It means that for $d=3$ and $r$ presumably within a certain vicinity of 0.3, the amplitude of
oscillations remains finite in the thermodynamic limit.
In our opinion, this is the first example of oscillatory behaviour in a microscopic model with
stochastic dynamics and without external periodic force.

Additional indication of anomalous behaviour can be seen in~\fig{f9} were we present the same data
as in~\fig{f8} but in the double-logarithmic scale.
All the data, except $d=3$ and $r=0.3$ approximately follow the solid line of slope 1/2 which
confirms the scaling $\sigma\sim \frac{1}{L^{d/2}}$.
However, for $d=3$ and $r=0.3$ one observes strong deviation from the expected scaling and
most likely (in agreement with~\fig{f8}) standard deviation will remain finite for
$L\rightarrow\infty$.
Let us also emphasize that simulations for $d=3$ and $r=0.3$ were rather extensive: we made runs of
$5\cdot 10^4$ Monte Carlo steps for systems of linear size up to $L=150$. 

To suggest some explanation of our results, let us first examine our model using the mean-field
approximation.
From the above stated dynamical rules, after neglecting some correlations, one can easily derive
the following mean-field equations~\cyt{LIPLIP}
\beqn
\frac{dx(t)}{dt}&=&rx(t)(1-x(t)^{2d})-(1-r)x(t)y(t), \label{2}\\
\frac{dy(t)}{dt}&=&(1-r)x(t)y(t)(1-y(t)^{2d})-(1-r)(1-x(t))y(t).
\eeqn{3}

These equations are very similar to~\eq{1} except that they contain some 'environmental capacity'
terms.
Although we did not succeed to solve~\eq{2}-\eq{3} analytically, these equations can be easily
solved numerically.
First, equating to zero the left-hand sides of~\eq{2}-\eq{3}, we obtain the so-called steady-state
equations and the solutions $x$ and $y$ of these equations for $d=3$ are shown in~\fig{f1}.

Numerical analysis indicates~\cyt{LIPLIP} that time-dependent solutions $x(t)$,
$y(t)$ of~\eq{2}-\eq{3} asymptotically (for infinite time) always approach the steady-state
solutions.
Since these mean-field equations include the 'environmental capacity' terms ($1-x(t)^{2d}$) and
($1-y(t)^{2d}$), the absence of limit-cycle solutions is an expected feature. 
However, for small $r$ an approach to the steady state proceeds through many oscillations and the
system resembles a weakly-damped two-dimensional oscillator.

In our opinion, this quasi-oscillatory behaviour suggests certain mechanism which can explain the
origin of such regular oscillations.
First, let us notice that noise, which is an intrinsic feature of the dynamics of our model, is
clearly neglected in the mean-field approximation~\eq{2}-\eq{3}.
In our opinion, when coupled to nonlinear oscillator~\eq{2}-\eq{3} this noise might, through
some sort of stochastic resonance, lead to the observed regular oscillations.

One indication of a resonatory mechanism is shown in~\fig{f6}, where for $r=0.3$ one can
see a second peak of $S(\omega)$ at approximately twice the frequency of the main peak.
Presumably, with more accurate calculations of the spectrum one could see also higher-order
harmonics.
Another indication is in our opinion the very shape of $\sigma$ as a function of $r$ for $d=3$
in~\fig{f7}.
Let us notice that $\sigma$ is a measure of fluctuations of $x(t)$ and thus might be regarded as a
response of our system to the noise.
From~\fig{f7} it is clear that the maximum of the response ($r\sim 0.3$) does not coincide with
the maximum of the noise (which most likely occurs at criticality i.e., at $r=r_{{\rm c}}\sim
0.05$) which is also a characteristic feature of resonatory systems.

An idea that random noise coupled to some low-dimensional autonomous system might lead to
oscillatory behaviour is not new. 
Some time ago Gang et al.~\cyt{GANG} studied a certain two-dimensional dynamical model with a point
attractor.
In its parameter space their model is located close to the region with limit-cycle attractor
and as a result some transient oscillations are observed.
Qualitatively their model is thus very similar to the system~\eq{2}-\eq{3}.
Gang et al. showed that when such a system is perturbed by a random noise, coherent oscillations
are observed, caused by some kind of stochastic resonance.
It might be interesting to examine the behaviour of the system~\eq{2}-\eq{3} subjected
to random noise.
However, since this system is only a low-dimensional approximation, it is by no means obvious that
it will correctly describe the behaviour of our microscopic model.
\section{Conclusions}
In the present paper we examined the time evolution of densities of preys and predators in a
certain lattice model.
As our main result we have found that for the three-dimensional case these densities might exhibit
very regular periodic oscillations.
We presented numerical evidence that the amplitude of these oscillations is nonzero even in the
thermodynamic limit.

Is it possible to suggest a certain general feature of our model which would be responsible for the
existence of such oscillations?
As far as the steady-state properties of the model are concerned the model has two phases: active
and absorbing.
Since the absorbing state is unique (all sites being occupied by preys), as expected, the
transition between them belongs to the directed-percolation universality class.
However, a closer look at the dynamics shows that there is yet another absorbing state in this
model: all sites being empty.
But this absorbing state is very unstable and the model almost never ends up in this state (a
single prey will invade the whole system in the absence of predators).
Although this absorbing state is irrelevant as far as the critical properties are concerned, this
state might, in our opinion, affect off-critical dynamical properties of our model:
First let us notice that empty sites are likely condidates for becoming occupied.
Thus, when large clusters of empty sites can be formed then large fluctuations of densities are
likely to happen too.
Such large clusters cannot form neither for large $r$ (almost all sites are occupied by preys and
predators) nor for small $r$ (for $r$ only slightly larger than $r_{{\rm c}}$ almost all sites are
occupied by preys) making the intermediate regime of $r$ the only possibility.
Let us also notice that this percolative argument explains the absence of oscillations for the 
$d=1$ case (there is no percolation in $d=1$ except for all sites being empty).
But to make this arguments more convincing it would be necessary to examine in details percolative
properties of our model.
Since the present model might be one of the simplest models exhibiting such oscillations, 
explaining its properties would be very desirable, especially because similar mechanism might be
responsible for oscillations in other spatially-extended systems.

Finally, we would like to make a very qualitative comparison of our results with experimental data
on oscillatory behaviour in prey-predator systems.
Although some oscillatory behaviour can be seen, these data (see e.g.,~\cyt{HALLAM}) clearly show
that these oscillations are very irregular.
In our opinion, qualitatively, these data are similar to our $d=2$ results rather than to $d=3$.
But this might not be very surprising since the 'world' of prey-predator systems for which these
data were collected, is basically two-dimensional.
Since some populations develop rather three-dimensional connections between individuals (e.g.,
fishes) it would be interesting to check whether oscillations in such populations are more regular.
\begin {thebibliography} {00}

\bib {NICOLIS} G.~Nicolis and I.~Prigogine, {\it Self-Organization in Nonequilibrium Systems}
(Wiley-Interscience, 1977).
Y.~Kuramoto, {\it Chemical Oscillations, Waves and Turbulence} (Springer, Berlin, 1984).

\bib {PRIG} I.~Prigogine and R.~Lefever, J.~Chem.~Phys.~{\bf 48}, 1695 (1968).

\bib {BARAS} F.~Baras, Phys.~Rev.~Lett.~{\bf 77}, 1398 (1996).
J.~Dethier, F.~Baras and M.~Malek Mansour, Europhys.~Lett.~{\bf 42}, 19 (1998).

\bib {TANNER} J.~T.~Tanner, Ecology {\bf 56}, 855 (1975).

\bib {LOTKA} A.~J.~Lotka Proc.~Natl.~Acad.~Sci.~USA {\bf 6}, 410 (1920).
V.~Volterra, {\it Lecon sur la Theorie Mathematique de la Lutte pour le via} (Gauthier-Villars,
Paris, 1931). 

\bib {HALLAM} T.~G.~Hallam, in {\it Mathematical Ecology}, T.~G.~Hallam and S.~A.~Levin eds.
(Springer-Verlag, Berlin, 1986).

\bib {SIGMUND} J.~Hofbauer and K.~Sigmund, {\it The Theory of Evolution and Dynamical Systems},
(Cambridge University Press, Cambridge, 1988).

\bib {GOEL} N.~S.~Goel, S.~C.~Maitra and E.~W.~Montroll, Rev.~Mod.~Phys.~{\bf 43}, 231 (1971).

\bib {LIPLIP} A.~Lipowski and D.~Lipowska, preprint (submitted to Physica A).

\bib {JANSSEN} H.~K.~Janssen, Z.~Phys.~B {\bf 42}, 151 (1981).
P.~Grassberger, Z.~Phys.~B {\bf 47}, 365 (1982).

\bib{GANG} H.~Gang, T.~Ditzinger, C.~Z.~Ning and H.~Haken, Phys.~Rev.~Lett.~{\bf 71}, 807 (1993).

\bib {HANGI} L.~Gammaitoni, P.~H\"{a}ngi, P.~Jung and F.~Marchesoni, Rev.~Mod.~Phys.~{\bf 70}, 223
(1998).

\end {thebibliography}
\begin {figure_captions}

\elem {f1} 
Steady-state densities of preys (dotted lines) and predators (dashed lines) for the one-
($\Box$) and three-dimensional ({\large +}) models as functions of $r$ as calculated using Monte
Carlo simulations~\cyt{LIPLIP}.
Mean-field results are shown as a solid line.

\elem {f2} 
Time evolution of $x(t)$ and $y(t)$ for the one-dimensional model and $r=0.6$.
Calculations were made for the linear size $L=2000$.

\elem {f3}
The power spectrum $S(\omega)$ for the one-dimensional model and $r=0.7$ ($\triangle$), 0.6
($\Diamond$), 0.55 ({\large +}) and 0.52 ($\Box$). Calculations are made for $L=10^4$.

\elem {time2} 
Time evolution of $x(t)$ (solid line) and $y(t)$ (dotted line) for the two-dimensional model and
$r=0.3$.
Calculations were made for the linear size $L=200$.

\elem {f4} The power spectrum $S(\omega)$ for the two-dimensional model and $r=0.6$ ($\star$), 0.5
($\triangle$), 0.4 ($\Box$), 0.3 ({\large +}) and 0.2 ($\Diamond$).
The increase of low-frequency part for decreasing $r$ is related with approaching the critical
point at $r\sim 0.11$ (critical slowing-down).

\elem {f5} 
Time evolution of $x(t)$ (solid line) and $y(t)$ (dotted line) for the three-dimensional model and
$r=0.3$.
Calculations were made for the linear size $L=30$.

\elem {f6}
The power spectrum $S(\omega)$ for the three-dimensional model and $r=0.5$ ($\triangle$), 0.4
($\Box$), 0.3 ({\large +}) and 0.2 ($\Diamond$).
Calculations are made for $L=30$.
The maximum value of $S(\omega)$ for $r=0.3$ is $S_{{\rm max}}\sim 0.12$.

\elem {f7} The standard deviation $\sigma$ as a function of $r$ for the one- ($\Box$), two-
($\Diamond$), and three-dimensional models ({\large +}).
For $d=1,2$ and 3 calculations were made for $L=30000, 150$ and 30, respectively.
For each value of $r$ we made runs of $2\cdot 10^4$ Monte Carlo steps.

\elem{f8} The standard deviation $\sigma$ as a function of $\frac{1}{L^{d/2}}$ for (a) $d=3$ and
$r=0.3 \ (\Diamond)$, (b) $d=3$ and $r=0.5 \ ({\large +})$, (c) $d=2$ and $r=0.3 \ (\Box )$, (d)
$d=1$ and $r=0.6 \ (\times)$.

\elem{f9} The standard deviation $\sigma$ as a function of $L^d$ in a double
logarithmic scale for (a) $d=3$ and $r=0.3 \ (\Diamond)$, (b) $d=3$ and $r=0.5 \ ({\large +})$, (c)
$d=2$ and $r=0.3 \ (\Box )$, (d) $d=1$ and $r=0.6 \ (\times)$.
The solid line has slope 1/2.

\end {figure_captions}
\end {document}
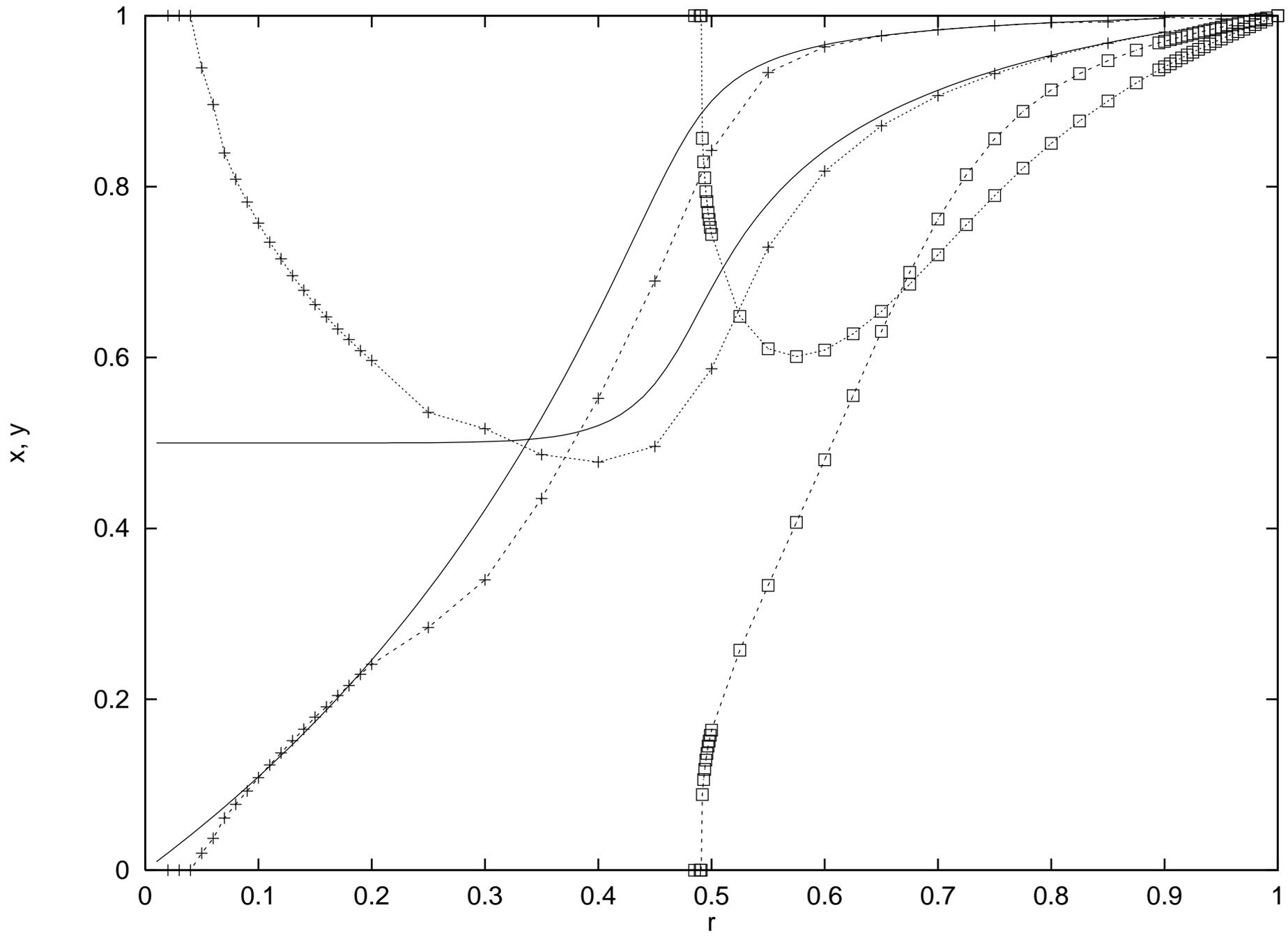
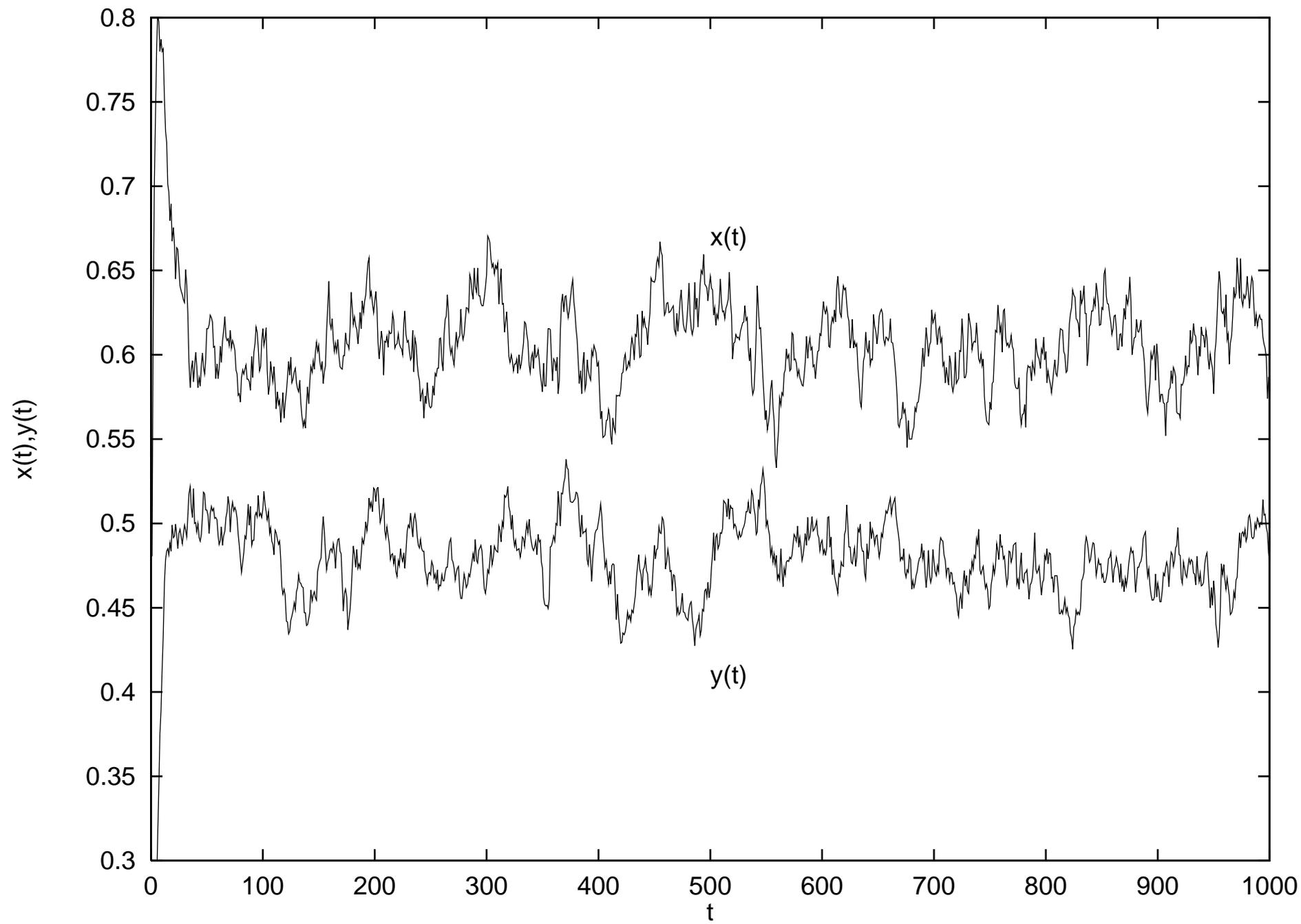
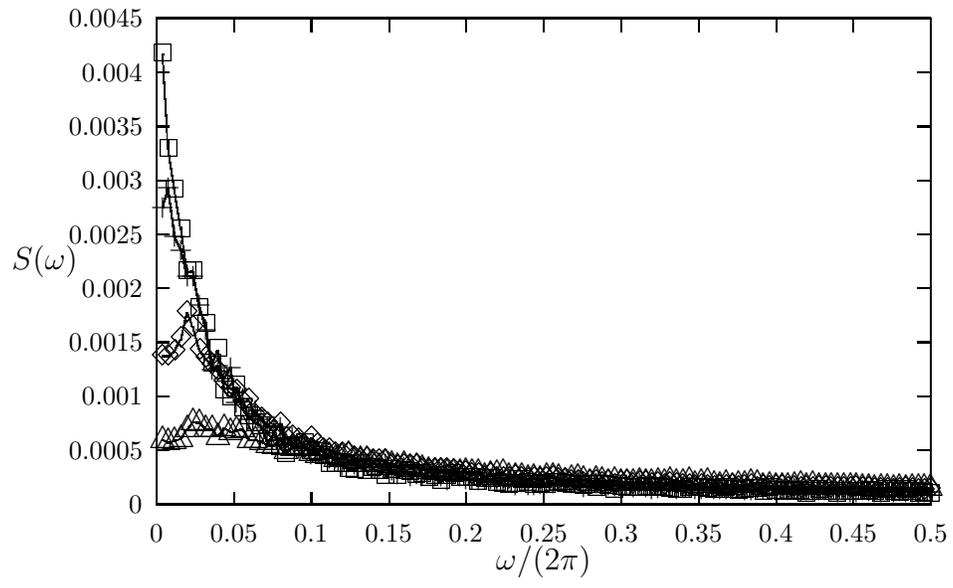
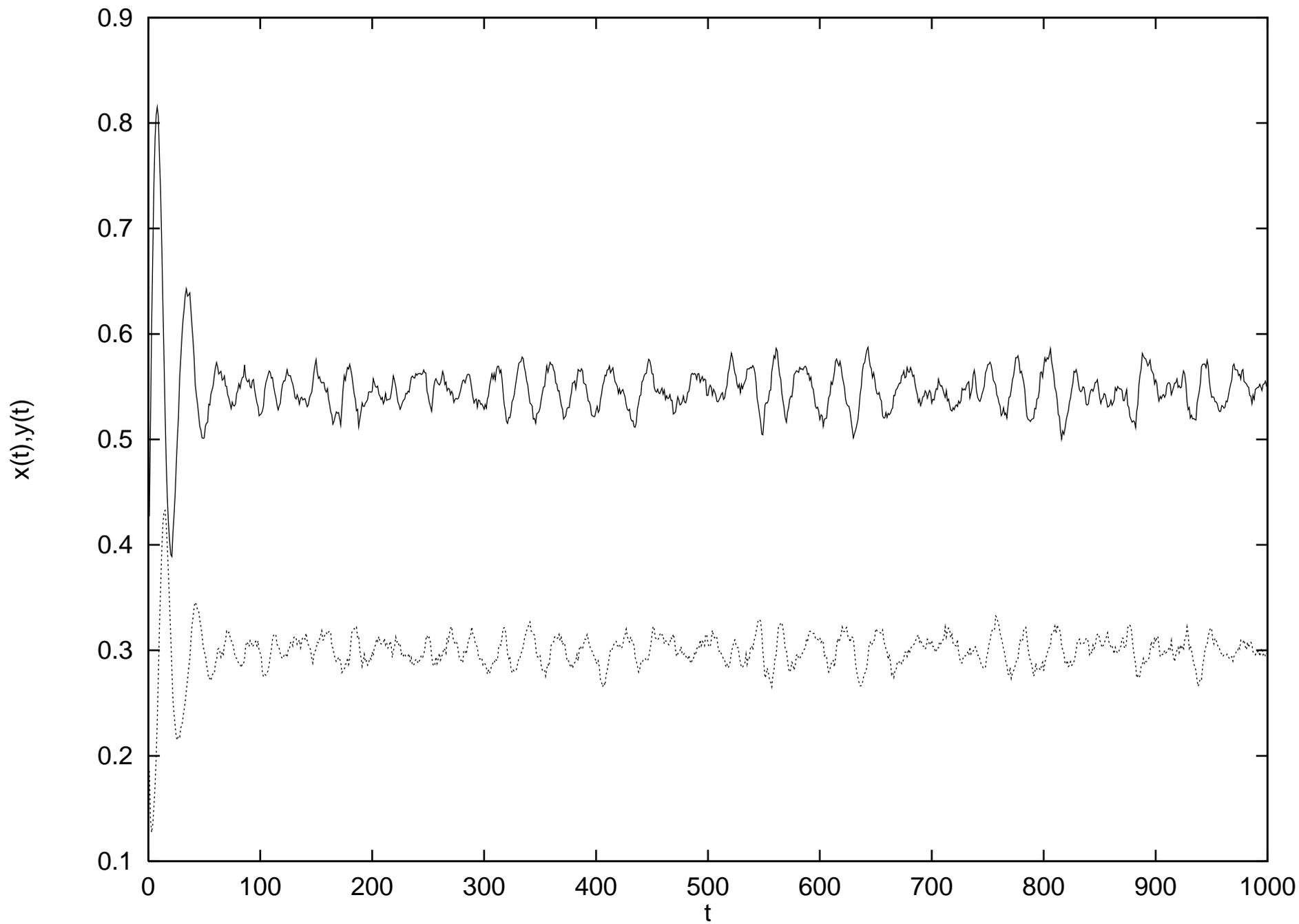
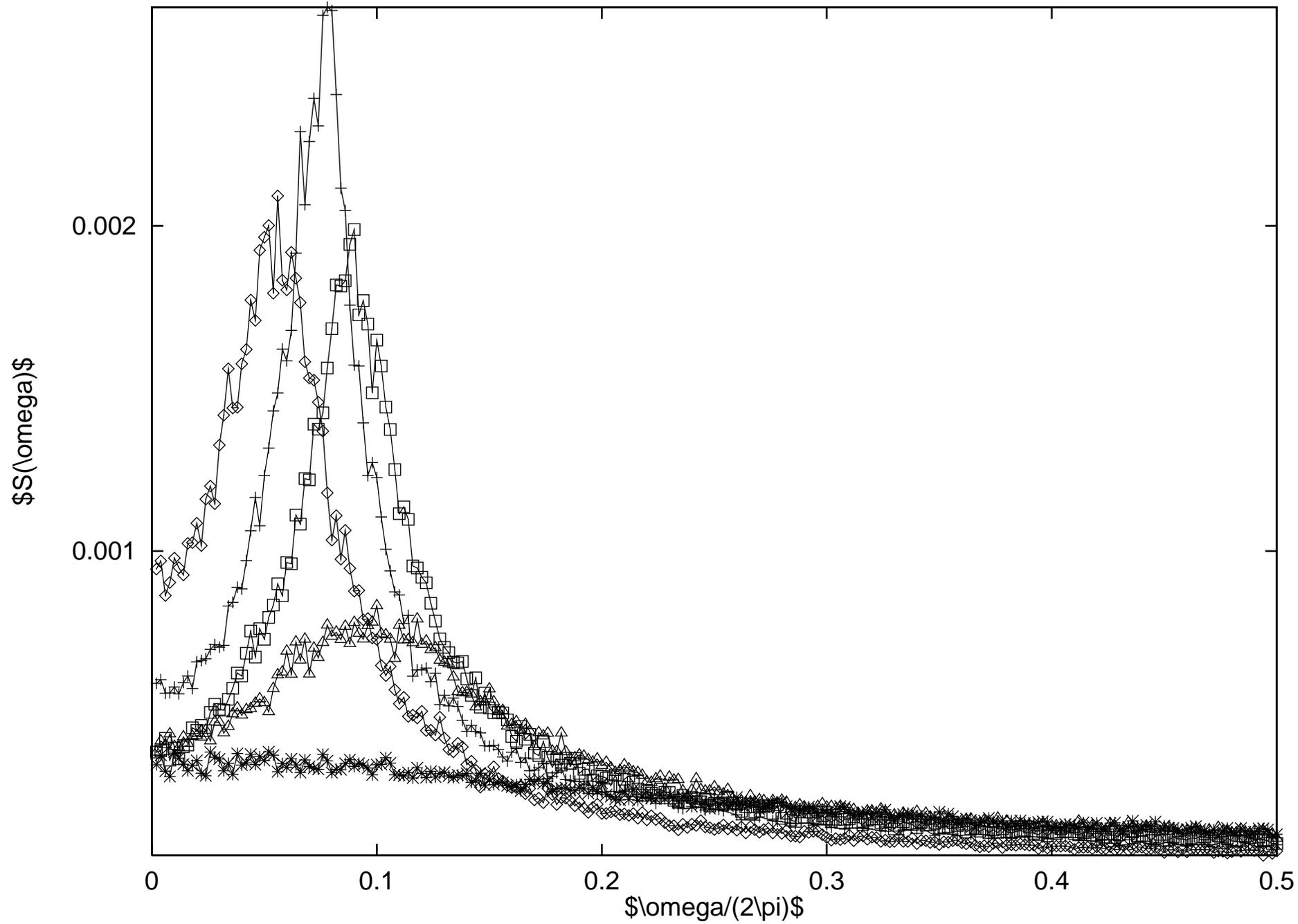
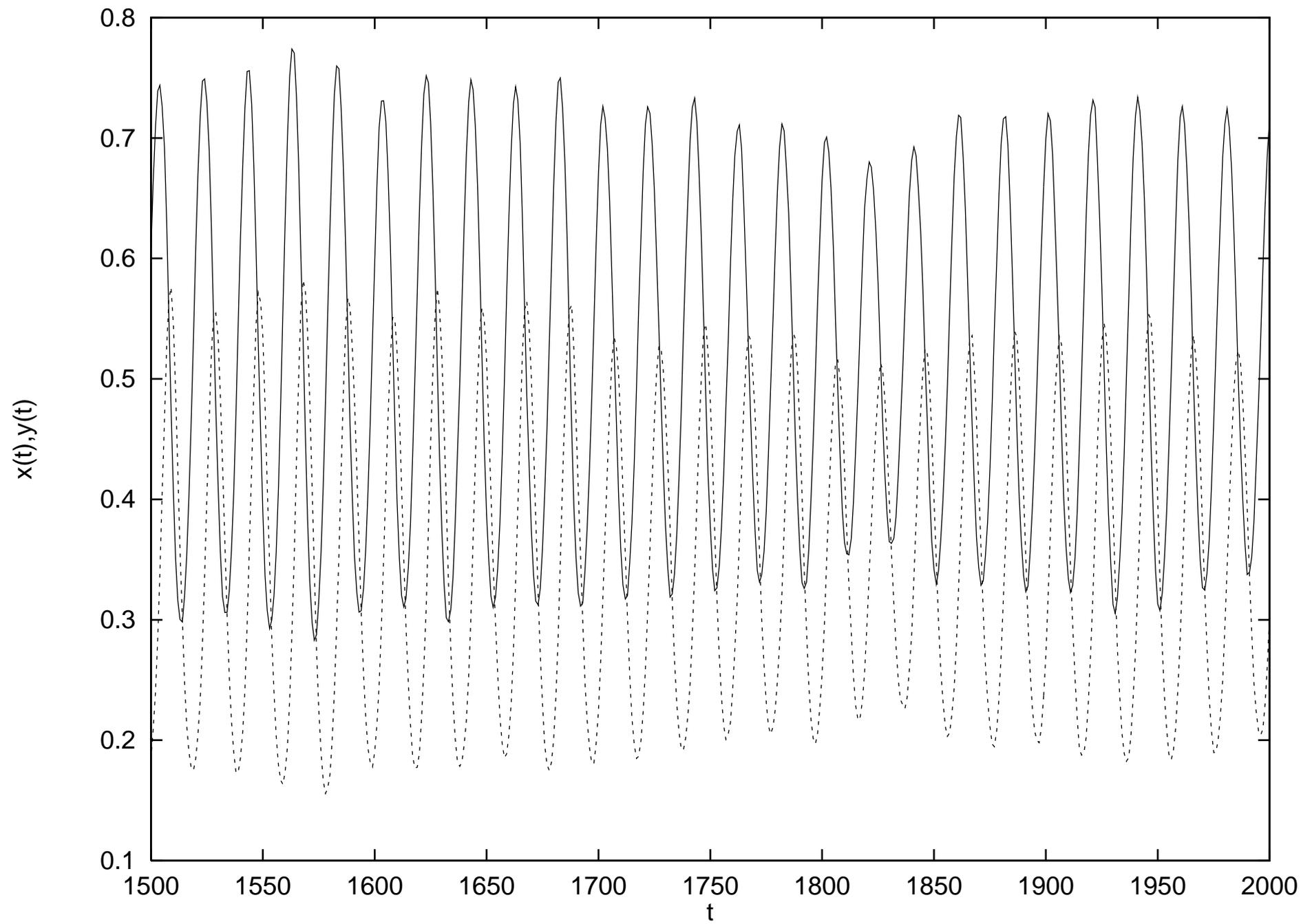
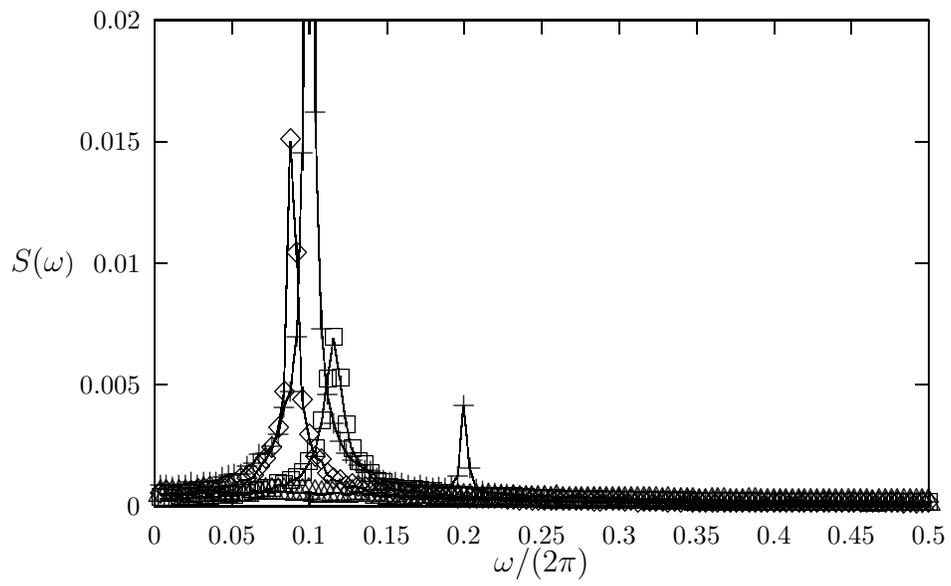
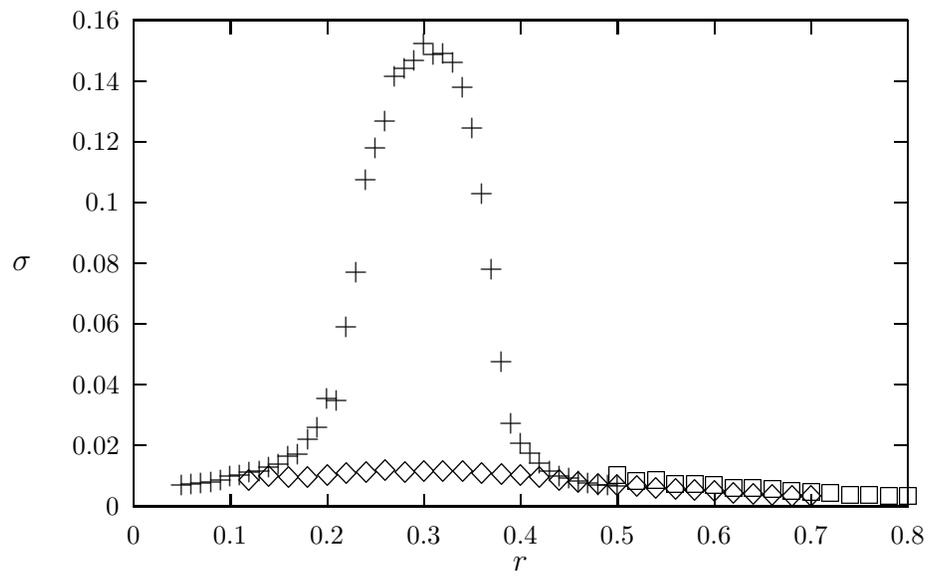
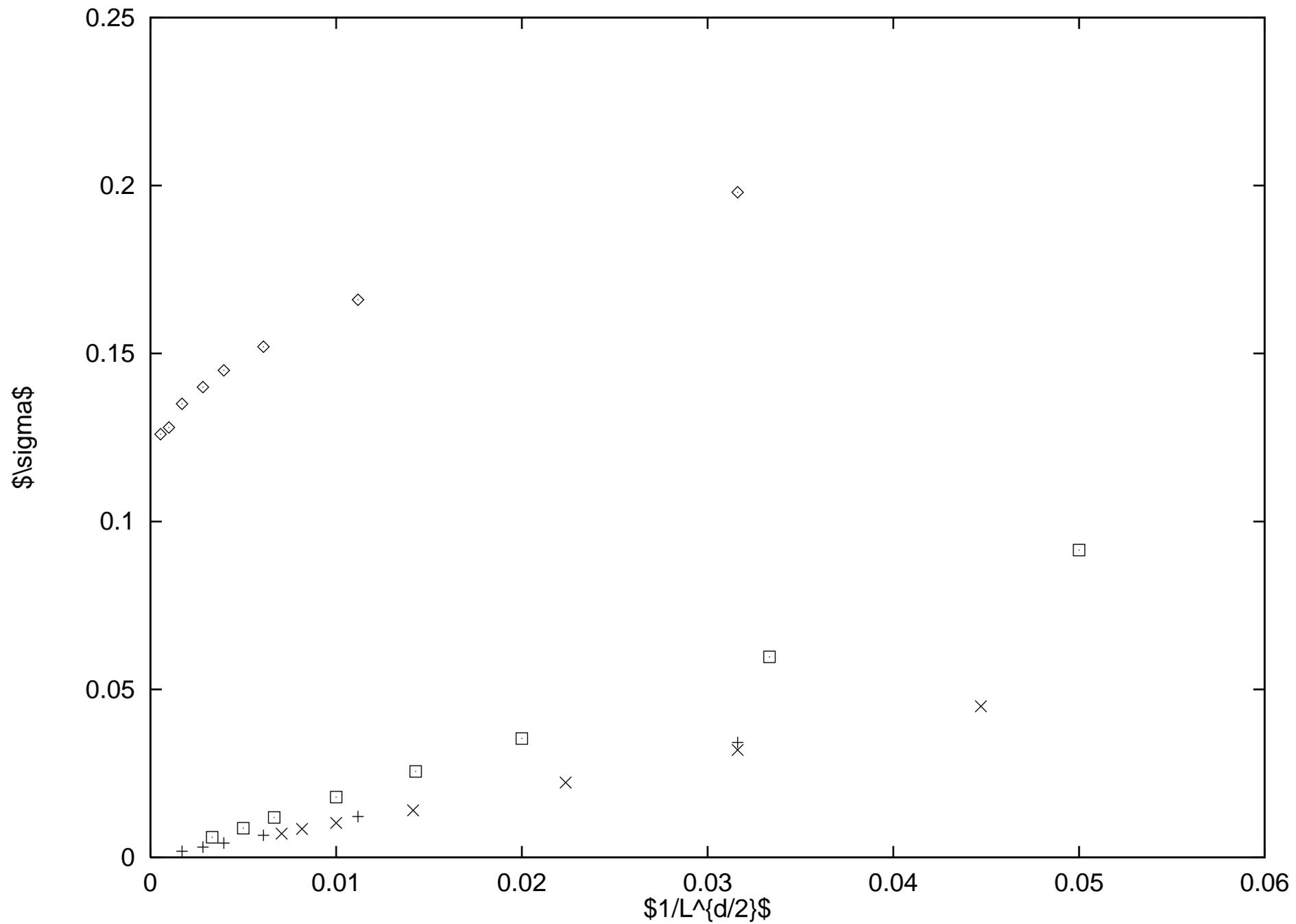
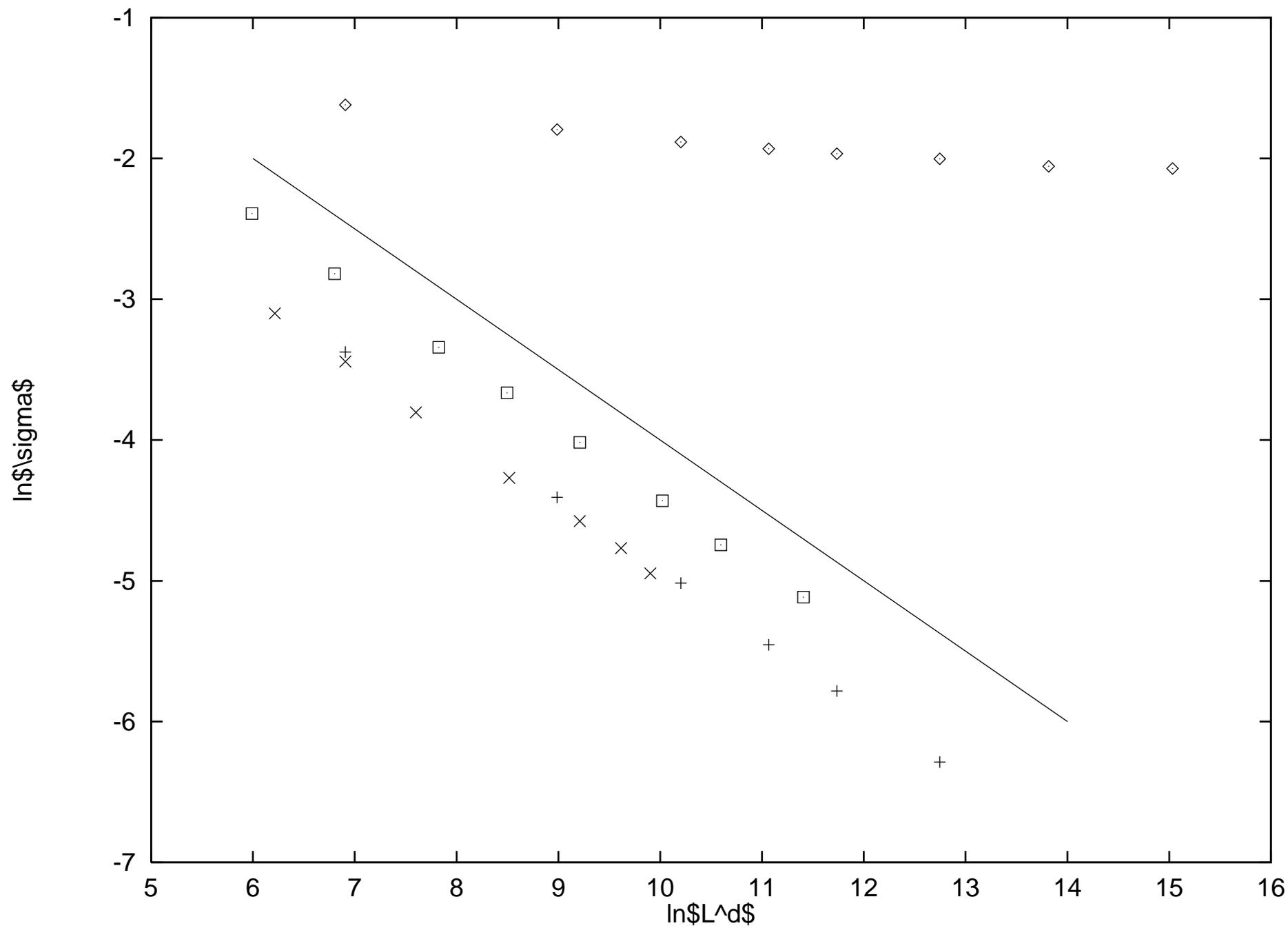
